\documentclass[12pt,epsf,amssymb,qsymbols]{article}
\usepackage{tabularx}
\usepackage{array}
\usepackage{graphics}
\usepackage{graphicx}
\usepackage{color}
\usepackage{epsfig}
\usepackage{amsmath}
\usepackage{amssymb}
\makeatletter

\usepackage{verbatim}

\newcommand{\msbar}{{\overline{\rm MS}}}

\newcommand{\nn}{\nonumber}
\newcommand{\bea}{\begin{eqnarray}}
\newcommand{\eea}{\end{eqnarray}}
\newcommand{\beq}{\begin{equation}}
\newcommand{\eeq}{\end{equation}}

\newcommand{\gev}{{\rm GeV}}

\newcommand{\pdir}{p\kern -5.2pt\raise 0.2ex\hbox {/}}
\newcommand{\vdir}{v\kern -5.75pt\raise 0.15ex\hbox {/}}
\newcommand{\kdir}{k\kern -5.75pt\raise 0.15ex\hbox {/}}
\newcommand{\epsdir}{\epsilon\kern -5.0pt\raise 0.15ex\hbox {/}}
\newcommand{\bvdir}{\bar{v}\kern -5.75pt\raise 0.15ex\hbox {/}}
\newcommand{\Ddir}{D\kern -7.75pt\raise 0.20ex\hbox {/}}
\newcommand{\ldir}{l\kern -5.0pt\raise 0.2ex\hbox{/}}
\newcommand{\varepsdir}{\varepsilon\kern -5.5pt\raise 0.15ex\hbox{/}}

\makeatother

\begin{document}
\begin{flushright}
\begin{tabular}{l}
{\tt }
\end{tabular}
\end{flushright}
\begin{center}
\vskip 1.cm\par
{\par\centering \Large \bf On the matrix elements of $\Delta B = 0$ operators}\\
\vskip .3cm\par
{\par\centering \Large \bf in the heavy meson decay widths}\\
\vskip 1.75cm
\par \scalebox{.89}{
\par\centering \large \sc Damir Be\'cirevi\'c$^a$, Svjetlana~Fajfer$^{b,c}$ and
Jernej~F.~Kamenik$^{b,d}$} {\par\centering \vskip 0.5 cm\par} {
\sl $^a$
Laboratoire de Physique Th\'eorique (Bat 210), \\ Universit\'e Paris
Sud, 91405 Orsay-Cedex, France.\\
\vspace{.25cm}
$^b$ J.~Stefan Institute, Jamova 39, P.O. Box 3000,\\
1001 Ljubljana, Slovenia.\\
%\vskip1.cm
\vspace{.25cm}
$^c$ Department of Physics, University of Ljubljana,\\
 Jadranska 19, 1000
Ljubljana, Slovenia.\\
\vspace{.25cm}
$^d$INFN, Laboratori Nazionali di Frascati, I-00044
Frascati, Italy.
%\vskip1.cm
}\\
{\vskip 0.25cm \par}
\end{center}

\vskip 0.25cm
\begin{abstract}
We determine the chiral corrections to the matrix elements of the $\Delta B=0$  four-quark operators which are relevant to the studies of the ratios of lifetimes of heavy-light mesons as well as to the power corrections to the inclusive semileptonic heavy-to-light decays. The chiral logarithmic corrections computed here can be combined with the forthcoming estimates of the corresponding matrix elements on the lattice to provide the reliable physics result of the well known bag-parameters $B_{1,2}$ and $\varepsilon_{1,2}$.
\end{abstract}
\vskip 0.2cm
\setcounter{page}{1}
\setcounter{footnote}{0}
\setcounter{equation}{0}
%%%%%%%%%%%%%%%%%%%%%%%%%%%%%%%%%%%%%%%%
%%%%%%%%%%%%%%%%%%%%%%%%%%%%%%%%%%%%%%%%
%%%%%%%%%%%%%%%%%%%%%%%%%%%%%%%%%%%%%%%%
\noindent

\renewcommand{\thefootnote}{\arabic{footnote}}
\vspace*{1.5cm}

%\newpage
\setcounter{footnote}{0}
%%%%%%%%%%%  Section 1

\noindent\underline{\bf 1. Phenomenological introduction}:  The matrix elements of $\Delta B=0$ operators 
enter several phenomenological studies of which the most important ones are the analyses of the 
spectra of inclusive semileptonic decays of heavy mesons and the lifetime ratios of heavy-light mesons.  

\noindent {\bf 1.1. Power correction to $\Gamma(B\to X_u e \nu)$}:  
Controlling the power corrections in the  spectra of inclusive semileptonic heavy to light 
decays has been --and still is-- an important obstacle when aiming at  the reliable extraction of 
the corresponding CKM parameters~\cite{Lange}. This is particularly important in the case of 
$\vert V_{ub}\vert$. In ref.~\cite{Voloshin} it has been shown that the $1/m_b^3$-corrections 
involve the matrix elements of dimension-6 four-quark operators of the flavour structure  $\Delta B=0$. 
More specifically 
\bea\label{eqq}
\Gamma(B\to X_u e \nu)_{1/m_b^3}&=&{G_F^2 m_b^5\over 192 \pi^3} \vert  V_{ub}\vert^2 
 \times {- 16 \pi^2\over m_b^3 } {1\over 2 m_B} \langle B\vert O_{V-A} - O_{S-P}\vert B\rangle  ,
\eea
where the matrix elements are conveniently expressed in terms of  bag parameters, 
$B_{1,2}$, as~\cite{Neubert}
\bea\label{bag1}
&&\langle B\vert O^u_{V-A}\vert B\rangle\equiv \langle B\vert \bar b \gamma_\mu (1- \gamma_{5} )  u\, 
\bar u \gamma^\mu (1- \gamma_{5} )  b\vert B\rangle =
f_{B}^2 m_{B}^2 B_1\,,\cr
&&\cr
&&\langle B\vert O^u_{S-P}\vert B\rangle\equiv \langle B\vert 
\bar b (1- \gamma_{5} )  u\, \bar u (1- \gamma_{5} )  b \vert B\rangle =
f_{B}^2 m_{B}^2 B_2\,,
\eea
with $f_B$ being the $B$-meson decay constant. Therefore what is actually needed in eq.~(\ref{eqq}) is the difference 
of bag parameters, $B_1-B_2$. The early estimates of $B_{1,2}$ in the framework of QCD sum rules in the static heavy quark limit were reported in 
ref.~\cite{qcdsr}, and recently extended to the full QCD case~\cite{gabbiani}. They were also computed on the lattice. In ref.~\cite{dipierro} the authors found that 
in the static heavy quark limit $B_1-B_2$ is zero, while the lattice study with the propagating heavy quark 
indicated that  $B_1-B_2$ can be quite different from zero~\cite{damir}. 
We are now in the era of ever better unquenched lattice QCD studies and a new 
lattice computation of the bag parameters of the $\Delta B=0$ operators is clearly desired. 
In recent years  it became evident that the control over the chiral extrapolation is essential in order to  
reduce the systematic uncertainties in the results of the lattice QCD studies.  In this  paper we provide 
the chiral corrections associated with the bag parameters computed in the static heavy quark limit. 
 
\noindent {\bf 1.2. $B_1 - B_2$ from $D$'s}:  
Before we turn to the question of chiral corrections, let us 
see if we can get some information about the size of $B_1-B_2$  from the available information on
  $D$-decays. The expression for the inclusive semileptonic decay 
width up to and including the terms $\propto 1/m_c^3$,  and neglecting the small contribution $\propto \vert V_{cd}\vert$ 
can be written as
\bea\label{eq0}
\Gamma(D\to X e \nu)&=&{G_F^2 m_c^5\over 192 \pi^3} \vert  V_{cs}\vert^2  \eta(z) 
 \left[
{1\over 2 m_D} \left( I_0(z) \langle D\vert \bar c c\vert D\rangle - {I_1(z)\over m_c^2}  \langle D
\vert \bar c g_s\sigma G c\vert D\rangle
\right)\right.\cr
&&\left. -{16 \pi^2\over 2 m_c^3 } f_D^2 m_D (B_1-B_2) \right]\, ,
\eea
where the phase space factor 
\bea
I_0(z) = (1-z^2)(1-8z+z^2)-12z^2\log z\,,\qquad I_1(z) = (1-z)^4\,,
\eea
and  $z=m_s^2/m_c^2$, while the numerical parameterisation of the $\alpha_s$-correction to the partonic decay width $\eta(z)$ reads~\cite{kim},
\bea
\eta(z) \approx 1 -{2\alpha_s\over 3\pi}\left[ \left(\pi^2-{31\over 4}\right) (1-z^{1/2})^2 +{3\over2} \right]\,.
\eea 
The equation of motion allows us to write
\bea
\bar c c = \bar c \vdir c + {1\over 2 m_c^2}\left( \bar c (iD_\perp )^2 c + \bar c \frac{g_s}{2}\sigma . G c \right)  +
{\cal O}(1/m_c^3)\,,
\eea
which in the standard notation~\footnote{
Also standard is the notation in terms of $\lambda_{1,2}$, the parameters 
measuring the kinetic and chromomagnetic energy of
the heavy quark inside a heavy-light system. The relation to $\mu_{G,\pi}^2$ is: 
$\lambda_1=-\mu_\pi^2$, and $\lambda_2=\mu_G^2/3$.}
\bea
\mu_\pi^2=-{1\over 2 m_D}\langle D\vert \bar c (i  D_\perp )^2 c\vert D\rangle\,,\qquad
\mu_G^2={1\over 2 m_D}\langle D\vert \bar c {g_s\over 2} \sigma . B c\vert D\rangle\,,
\eea
can be written as
\bea
{1\over 2 m_D}\langle D\vert \bar c c\vert D\rangle = 1 -{ \mu_\pi^2 - \mu_G^2 \over 2 m_c^2}\,.
\eea
Finally  eq.~(\ref{eq0}) becomes
\bea\label{eqA}
\Gamma(D\to X e \nu)&=&{G_F^2 m_c^5\over 192 \pi^3} \vert  V_{cs}\vert^2 I_0(z)\eta(z)
 \left\{
1 + {1\over 2 m_c^2}\left[ \mu_\pi^2 - {\left( 1 - 4 I_1(z)/I_0(z) \right)\mu_G^2} \right]\right\}\cr
&&\cr
&& -{G_F^2 m_c^2\over 12 \pi} \vert  V_{cs}\vert^2 \eta(z)  f_D^2 m_D \left(B_1-B_2\right) \,.
\eea
Clearly the heavy quark expansion applied to the decay of charmed mesons is expected to converge much slower
than in the case of $B$-mesons. It is however interesting to use the available information on charmed modes to  
bound the $B_1-B_2$ value. 
Concerning the parameters appearing in the first line of eq.~(\ref{eqA}) we can use
$\mu_G^2=(3/4)[m_{D^\ast}^2-m_D^2]= 0.41~\gev^2$, while the value of $\mu_\pi^2$ is still somewhat vague. 
Recent experimental fits to the moments of the semileptonic $b\to c$ decay spectrum~\cite{belle,babar,delphi}  
quote $m_c \approx 1.1(1)~\gev$ and $\mu_\pi^2 \approx 0.5(1)~\gev^2$
in the so-called kinetic scheme~\cite{uraltsev} and at $\mu=1$~GeV. When converted to the $\msbar$ scheme, the charm quark mass is $m_c(m_c)=1.2(1)$~GeV, consistent with the estimates based on the lattice QCD simulations 
$m_c(m_c)=1.32(3)$~GeV, and $1.30(3)$~GeV~\cite{MC-latt}, as well as with the recent  
QCD sum rule study  $m_c(m_c)=1.29(1)$~GeV~\cite{MC-QCDSR}.
If,  in addition, we take  $f_D=208(4)$~MeV~\cite{follana}, $\tau_{D^\pm}=1.040(7)$~ps,  $\tau_{D^0}=0.410(15)$~ps~\cite{PDG}, and the recently measured semileptonic 
branching fractions~\cite{CLEOc}:~\footnote{
Very recently the other charm factory (BES) presented similar results but with the final muon instead of electron~\cite{bes}. 
Their values are fully consistent with those given in eq.~(\ref{ccleo}), measured by CLEOc, but the error bars are an order of magnitude larger.} 
\bea\label{ccleo}
B(D^+ \to  X  e\nu) = (16.13 \pm 0.20 \pm 0.33)\%\,,~~  B (D^0 \to X e\nu ) = (6.46\pm0.17\pm0.13)\%\,,
\eea
then we get that the terms including $1/m_c^2$-corrections saturate the 
experimental value for the branching ratio to about $85 \%$.  If we now assume 
the $1/m_c^3$ term fully saturates the rate, we get $B_2(m_c)-B_1(m_c) \approx 0.04$. When  evolving 
those bag parameters to the $m_b$-scale~\cite{Neubert,roma3}, the difference is further increased by about $30\%$, leading us to $B_2(m_b)-B_1(m_b) \approx 0.05$. 
After plugging that number in eq.~(\ref{eqq}), and by taking $f_B=0.2$~GeV, we get
\bea 
B(B\to X_u e \nu)_{1/m_b^3}&<& 0.04 \times \vert V_{ub}\vert^2   ,
\eea
which is  ({\sl comfortably}) a very small number. Of course this exercise is only a speculation, while for the reliable estimate of $B(B\to X_u e \nu)_{1/m_b^3}$ a direct  non-perturbative method should be employed to compute the matrix elements~(\ref{bag1}).

\noindent{\bf 1.3. $B$-meson lifetimes}:
When studying the $B$-meson lifetimes, due to the nonleptonic decay modes two more operators enter the game
\bea\label{bag2}
&&\langle B\vert T^q_{V-A}\vert B\rangle\equiv \langle B_q\vert \bar b{ \lambda^A\over 2} \gamma_\mu (1- \gamma_{5} )  q\, 
\bar q { \lambda^A\over 2} \gamma^\mu (1- \gamma_{5} )  b\vert B_q\rangle =
f_{B}^2 m_{B}^2 \varepsilon_1\,,\cr
&&\cr
&&\langle B\vert T^q_{V-A}\vert B\rangle\equiv \langle B_q\vert \bar b{ \lambda^A\over 2} (1- \gamma_{5} )  q\, 
\bar q { \lambda^A\over 2}  (1- \gamma_{5} )  b\vert B_q\rangle =
f_{B}^2 m_{B}^2 \varepsilon_2\,,
\eea
where ${ \lambda^A}$ are the Gell-Mann matrices. The bag parameters in 
eqs.~(\ref{bag1},\ref{bag2}) were first introduced in this form in ref.~\cite{Neubert}
 in  which the authors studied the lifetime difference of hadrons containing one valence $b$-quark. 
Voloshin, however, realised in ref.~\cite{Voloshin} that the light 
flavour content of the operators can be different from  the light valence quark of the $B$-meson. 
Those are the (in)famous ``eye-contractions" which are extremely difficult to study non-perturbatively. 
In addition, an extra penguin operator contribution was singled out in  ref.~\cite{roma3}. The contribution 
of the non-valence and the penguin operators are expected to be negligible in the case of 
the meson lifetime ratios due to the light flavour symmetry of the spectator quark. That argument however does not apply to 
the power correction to the semileptonic decays~(\ref{eq0}).
In terms of bag parameters and neglecting the eye-contractions as well as  the penguin operator contributions, but including 
the NLO QCD-corrections to the Wilson coefficients computed in refs.~\cite{roma3,lenz}, 
the  master formulas for the lifetime ratios of $B$-mesons read
\bea\label{lifediff}
{\tau(B^\pm)\over \tau(B_d)} &=& 1 + 0.07(2)\times  B_1 + 0.011(3) \times B_2 - 0.7(2) \times  \varepsilon_1 + 0.18(5) \times \varepsilon_2\,,\cr
{\tau(B_s)\over \tau(B_d)} &=& 1 + 0.007(2)\times  [ B_1^s - B_1] - 0.009(2)\times  [B_2^s -0.9 B_2]  \cr 
&&+ 0.15(4)\times  [
 \varepsilon_1^s-1.1\varepsilon_1] - 0.18(5) \times [\varepsilon_2^s-0.9\varepsilon_1]\,,
\eea
where the superscript $``s"$ has been used to distinguish the bag parameters for the case of valence strange quark. 
In this case it is even more important to have a good handle on the $\varepsilon_{1,2}$ parameters whose impact is
enhanced by the size of the Wilson coefficients (numerical values of which are displayed above).

\vspace*{.75cm}

\noindent\underline{\bf 2. Chiral corrections}: 
One of the main problems in relating  the bag parameters $B_{1,2}$ and $\varepsilon_{1,2}$ computed on 
the lattice to the physical bag parameters  is the necessity to perform the chiral extrapolation of matrix elements 
computed with the light quark masses directly accessible on the lattice ($1> m_q/m_s \gtrsim 1/4$) 
down to the physical limit ($ m_q/m_s \approx 1/25$).~\footnote{Here and in the following $m_q\equiv m_u=m_d$.} The expressions derived in chiral perturbation theory 
provide an important guidance in that respect. 
The chiral corrections to the matrix elements of the whole basis of four-quark  $\Delta B=2$ operators were recently  computed 
in refs.~\cite{lin,our}. We showed in ref.~\cite{our} that the validity of the formulas derived in heavy meson chiral perturbation theory (HMChPT) may be questionable for the quarks 
not lighter than  about the third of the strange quark mass, because of the nearness of the scalar heavy-light mesons 
(or more precisely, of the heavy-light $(\frac{1}{2})^+$-doublet). In other words, unless one wants to deal with a very large number of low energy constants,  the adequate HMChPT  expressions are only those with $N_f=2$ light quark flavours (i.e. with the pion loops only). 
In this paper we do not return to that issue. Instead we focus on the chiral corrections to the bag parameters of  the $\Delta B=0$ operators introduced above.

\noindent{\bf 2.1. Framework}: 
As in ref.~\cite{our}, in the present paper  we work in the static heavy quark limit 
and use the HMChPT lagrangian already described in 
detail in ref.~\cite{our}. We chose a basis of operators
\bea
 \label{baseS}
{\phantom{{l}}}\raisebox{-.16cm}{\phantom{{j}}} O_{V-A}
&=& \ \bar b \gamma_\mu (1-\gamma_5)  q_L \,
 \bar q_L  \gamma^\mu (1-\gamma_5) b \,  ,
  \nonumber \\
{\phantom{{l}}}\raisebox{-.16cm}{\phantom{{j}}} T_{V-A}
&=& \ \bar b \gamma_\mu (1-\gamma_5) t^A q_L \,
 \bar q_L t^A \gamma^\mu (1-\gamma_5)  b \,  ,
  \nonumber \\
{\phantom{{l}}}\raisebox{-.16cm}{\phantom{{j}}} O_{S-P}
&=& \ \bar b (1-\gamma_5)
 q_L \,
 \bar q_L  (1+\gamma_5)  b \,  ,  \nonumber \\
{\phantom{{l}}}\raisebox{-.16cm}{\phantom{{j}}} T_{S-P}
&=& \ \bar b (1-\gamma_5)
t^A q_L \,
 \bar q_L t^A (1+\gamma_5)  b \,  .  \nonumber
 \eea
The heavy quark spin ($S$) and the chiral symmetry ($U_L, U_R$) transformations 
act on the heavy and light quark respectively like,  $b\to S b$ (i.e., $\gamma_0 b = b$), and  
 $q_{L,R} \to U_{L,R} q_{L,R}$. Since the colour structure ({\sl short distance}) does not 
influence the chiral logarithms ({\sl long distance})~\cite{gio},  we need to consider only 
two of the above operators which we choose to be  $ 
O_{V-A}$ and $ O_{S-P}$. In HMChPT we need the bosonised forms of these operators, which 
are built up from the heavy-light $(\frac{1}{2})^-$-doublet fields, 
$H_q(v) = {1+\vdir \over 2}\left[ P_{q \mu}^{ \ast
}(v)\gamma^\mu - P_q(v) \gamma_5\right]_q$, 
and the pseudo-Goldstone fields, $\Sigma =
\exp(2 i \phi / f)$, where $\phi$ is the usual matrix of pseudo-Goldstone bosons.  Under the heavy
quark and chiral symmetry the field  $H_q(v)$ transforms as $H_q \to S H_{q'}
U_{q'q}^{\dagger}$, while  $\Sigma$  transforms as $\Sigma \to U_L \Sigma
U_R^{\dagger}$. The standard procedure then consists  in introducing  
$\xi = \sqrt \Sigma = \exp(i \phi /f)$, which transforms as $\xi \to U_L \xi U^{\dagger} = U \xi
U_R^{\dagger}$.  By  a simple chiral and heavy quark spurion analysis of the
operators $O_{V-A}$,  $O_{S-P}$ we then obtain their most general bosonized
form within HMChPT, namely 
\bea\label{bosbase} 
O^{q}_{V-A} &=& \sum_X \tau_{1X} {\rm Tr} \left[ (\xi \overline H)_q
\gamma_{\mu}
(1-\gamma_5) X \right] {\rm Tr} \left[ X \gamma^{\mu} (1-\gamma_5) (H \xi^{\dagger})_{q} \right] \nn \\
&&+  \sum_{X,q'} \delta_{1X} {\rm Tr} \left[ \overline
H_{q'} \gamma_{\mu} (1-\gamma_5) X \right] {\rm Tr} \left[  X \gamma^{\mu} (1-\gamma_5) H_{q'} \right]+ {\rm c.t.}\,,\nn \\ &&\hfill \nn \\
 O^{q}_{S-P} &=& \sum_X \tau_{2X} {\rm Tr} \left[ (\xi
\overline H)_q (1-\gamma_5) X \right] {\rm Tr} \left[ X (1+\gamma_5)  ( H
\xi^{\dagger})_{q} \right]  \nn\\
&&+ \sum_{X,q'} \delta_{2X} {\rm Tr} \left[ \overline
H_{q'} (1-\gamma_5) X \right] {\rm Tr} \left[  X (1+\gamma_5) H_{ q'} \right]+ {\rm c.t.}\,,
\eea where ``c.t." stands for counterterms, and  $X\in\{ 1, \gamma_5, \gamma_{\nu}, \gamma_{\nu}\gamma_5,
\sigma_{\nu\rho}\}$.  The contraction of Lorentz indices and HQET parity conservation requires 
the same $X$ to appear in both traces in the products. Any insertion of $\vdir $ can be
absorbed via equation of motion, $\vdir  H(v) = H(v)$, while any nonfactorizable
contribution with a single trace over Dirac matrices can be reduced
to the  form written in~(\ref{bosbase}) by using the $4\times 4$ matrix identity 
\bea 
4\ {\rm Tr}(AB) &=& {\rm Tr}(A){\rm Tr}(B) + {\rm Tr}(\gamma_5 A){\rm
Tr}(\gamma_5 B) +
{\rm Tr}(A\gamma_{\mu}){\rm Tr}(\gamma^{\mu}B) \nn \\
&&+{\rm Tr}(A\gamma_{\mu}\gamma_5){\rm Tr}(\gamma_5\gamma^{\mu}B) +
\frac{1}{2}{\rm Tr}(A\sigma_{\mu\nu}){\rm Tr}(\sigma^{\mu\nu}B). 
\eea
Note that these matrix identities also ensure invariance of the
bosonized operators under Fierz transformations, as can be readily
checked by bosonizing the Fierz transformed quark level operators,
which involve formally different spurion fields.

An important comment is also that the second lines in the operators 
written in eq.~(\ref{bosbase})  stand for the corresponding 
``eye-contractions".  The sum over $q'$ runs over all light quark flavours. Once 
saturated by the external heavy-light meson states of the light flavour $q$ only the eye-contraction 
with  $q=q'$ will contribute.  
However  at  one loop  in HMChPT these contractions will not produce any chiral logarithmic 
correction to the matrix  elements of the above operators. Their effect can show up at two or more loops though.

After calculating the above traces, and by retaining the pseudo-Goldstone
fields $\phi$ up to quadratic order, we obtain:
\bea\label{bosbase:expanded}
  O^{ q}_{V-A} &=& 4\widehat \tau_1 (P_{q'\mu}^\dagger P_{ q'}^{\mu} + P_{q'}^{\dagger} P_{ q'}) \left[\delta_{q' q}  \left(1+ \frac{\hat\delta_1}{\hat \tau_1} + \frac{i}{f} (\phi_{q' q}  -   \phi_{  q'  q})\right) \right.\nn\\
&& \left.  + \frac{1}{2 f^2}\biggl (2 \phi_{q' q} \phi_{q'  q} - \delta_{q' q} (\phi.\phi)_{ q'  q} - (\phi.\phi)_{q' q} \delta_{ q'  q}\biggr) \right] + \ldots \,,\nn \\
 &&\hfill \nn \\
  O^{q}_{S-P} &=&  4(\widehat \tau^*_2 P_{q'\mu}^\dagger P_{ q'}^{\mu} +  \widehat \tau_2 P_{q'}^{\dagger} P_{ q'}) \left[\delta_{q' q}  \left(1+ \frac{\hat\delta_2^{(\ast)}}{\hat \tau_2^{(\ast)}} + \frac{i}{f} (\phi_{q' q}  -   \phi_{  q'  q})\right) \right.\nn\\
&& \left.  + \frac{1}{2 f^2}\biggl (2 \phi_{q' q} \phi_{q'  q} - \delta_{q' q} (\phi.\phi)_{ q'  q} - (\phi.\phi)_{q' q} \delta_{ q'  q}\biggr) \right] + \ldots  \,,
 \eea
where, for simplicity,  we do not display the counterterms and we used
\bea\label{betas} \widehat \tau_1^{(*)} &=& \tau_1 +
\tau_{1\gamma_5} - 4 (\tau_{1\gamma_{\nu}} +
\tau_{1\gamma_{\nu}\gamma_5}) - 12 \tau_{1\sigma_{\nu\rho}},\nn\\
\widehat \tau_2 &=& \tau_2 - \tau_{2\gamma_5} - \tau_{2\gamma_{\nu}}
+ \tau_{2\gamma_{\nu}\gamma_5} + \bar \tau_2- \bar \tau_{2\gamma_5}
- \bar \tau_{2\gamma_{\nu}} + \bar
\tau_{2\gamma_{\nu}\gamma_5},\nn\\
\widehat \tau_2^* &=& \tau_{2\gamma_{\nu}} -
\tau_{2\gamma_{\nu}\gamma_5} + \bar \tau_{2\gamma_{\nu}} - \bar
\tau_{2\gamma_{\nu}\gamma_5}. \eea
Similarly $\hat \delta_{1,2}^{(\ast)}$ stand for the combinations of $\delta_{1X,2X}$ couplings appearing in eq.~(\ref{bosbase}).

\noindent{\bf 2.2. Chiral loop corrections}
%%%%%%%%%%%%%%%%%%%%%%%%%%%%%%%%%%%%%%%%%%%%%%%%%%%%%%%%%%%
\begin{figure}
\begin{center}
\epsfxsize14.6cm\epsffile{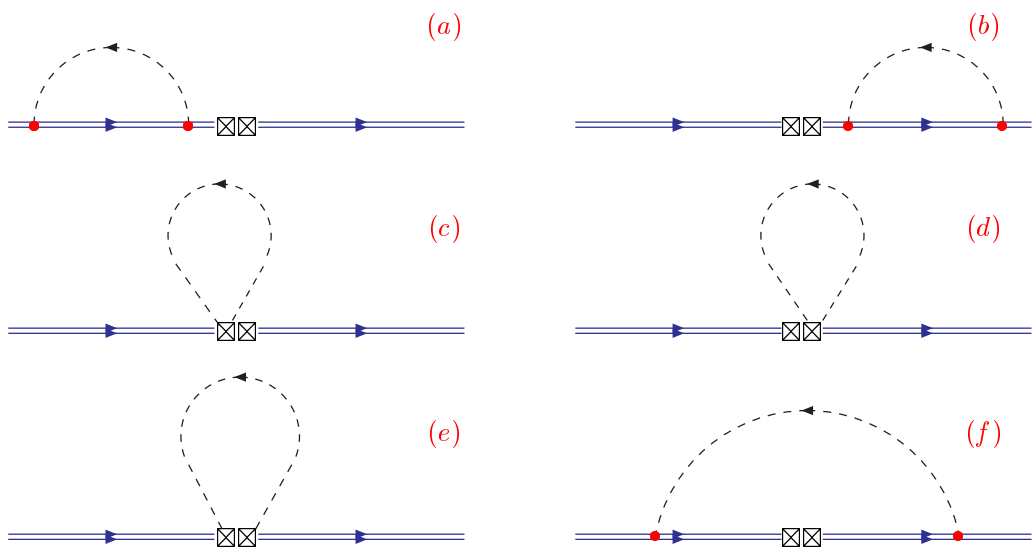}   \\
%%%%%%%%%%%%%%%%%%%%%%%%%%%%%%%%%%%%%%%%%%%%%%%%%%%%%%%%%%%%%%%%%%
\caption{\label{fig1}{\small \sl The graphs giving the non-zero contribution to the NLO chiral corrections 
to the matrix elements of the $\Delta B=0$ operators  discussed in this paper. The double lines correspond to the 
heavy-light mesons, and the dashed ones to the pseudo-Goldstone bosons. $\Delta B=0$ operators are denoted 
by  ``$\boxtimes\!\boxtimes$", while the strong vertices coming from the HMChPT lagrangian 
are denoted by the full dots, ``{\color{red}{$\bullet$}}".  }}
%%%%%%%%%%%%%%%%%%%%%%%%%%%%%%%%%%%%%%%%%%%%%%%%%%%%%%%%%%%%%%%%%%
\end{center}
\end{figure}
%%%%%%%%%%%%%%%%%%%%%%%%%%%%%%%%%%%%%%%%%%%%%%%%%%%%%%%%%%%
The computation of the chiral loop corrections to our operators is by now standard. 
It involves $6$ diagrams which are  shown in fig.~\ref{fig1}.  Four graphs are factorizable 
and  two are not. Of factorizable diagrams we have the self energy contributions [$(a)$ and $(b) $ 
in fig.~\ref{fig1}] which give rise to the wave function renormalisation corrections ($\delta Z_q$) which can be found in e.g.~\cite{our}, 
and two tadpole graphs which represent the loop corrections to the weak currents composing 
the four-quark $\Delta B=0$ operator  [$(c)$ and $(d) $ in fig.~\ref{fig1}].  
Finally there are two nonfactorizable graphs: tadpole  [$(e)$],  and ``sunset"  [$(f)$].  
An alerted reader may notice the absence of the mixed terms, i.e. the ones involving an
exchange of a pseudo-Goldstone boson between the weak operator and the
HMChPT interaction Lagrangian. Those contributions drop out due to heavy vector meson
transversality ($v \cdot \varepsilon_{P^*}=0$).
Compared to the situation we encountered in the computation of  the chiral loop corrections to the matrix elements 
of   $\Delta B=2$ operators~\cite{our}, in the present situation  the sum of 
diagrams $(c)$, $(d)$ and $(e) $ vanishes. 

The resulting expressions read 
\bea f_q^2 B^{q }_{1} &=& (f_q^2 B_1)^{\rm Tree}
 \left\{
1 + \delta Z_q   - \frac{1}{2f^2}[ 6
g^2 t^i t^{i\dagger} + X_1 (t^{i\dagger} t^{i\dagger} + t^i t^{i} -
2 t^i t^{i\dagger}) ]_{q q}
I_0(m_i) + \mathrm{c.t.}\right\}\,,\nn\\
f_q^2 B^{q }_{2} &=& (f_q^2 B_2)^{\rm Tree}
 \left\{
1 +  \delta Z_{\bar q} - \frac{1}{2f^2}[ 6
Y_2 g^2 t^i t^{i\dagger} + X_2 (t^{i\dagger} t^{i\dagger} + t^i
t^{i} - 2 t^i t^{i\dagger}) ]_{q  q} I_0(m_i) +
\mathrm{c.t.}\right\}\,,\nn\\ \eea where $I_0(m_i)= (m_i/4\pi)^2
\log (m_i^2/\mu^2)$, $Y_{2} = (B^*_{2}/ B_{2})^{\rm Tree}$, $X_{i} =
(\widehat \tau_{i}/ B_{i})^{\rm Tree}\approx (1-\widehat
\delta_{i}/\widehat \tau_{i})$, and $t^i$ are the $SU(N)$ generator
matrices. Summation over  ``$i$" in the above expressions  is understood. 

\vspace*{.75cm}

\noindent\underline{\bf 3. Results}: On  the basis of the expressions derived in the previous section we now discuss our results. We will first give the explicit formulas for  the chiral corrections that might be particularly useful to the lattice practitioners. We will then make some important assumptions which will allow us to infer a few  phenomenological implications.

\noindent{\bf 3.1. Message relevant to the lattice QCD studies}:
In our previous paper we showed that due to the nearness of the $\left(\frac{1}{2}\right)^+$-doublet 
of the heavy-light mesons, only the pion loop contributions are a safe prediction of this 
(HMChPT) approach which then can be used to guide  the chiral extrapolations of 
the heavy-light meson quantities computed on the lattice. This was shown to be the case 
for the decay constants, Standard Model and SUSY bag parameters parameterising the
matrix elements of the  $\Delta B=2$ operators~\cite{our}, pionic couplings $g$, $\widetilde g$ and $h$~\cite{svje}, as 
well as for the Isgur-Wise functions~\cite{jern}. The same holds true in this case. 
Therefore the relevant expressions to be used in the lattice  extrapolations in the light quark mass 
are those  derived in $SU(2)$-theory and they read:
\bea f_q^2 B_1 &=&\alpha^2 B_1^{\rm Tree}
 \left[
1 - {9 g^2  \over (4\pi f)^2} m_\pi^2\log{m_\pi^2\over
\mu^2} + o_1(\mu) m_\pi^2 \right]\,,\nn\\
f_q^2 B_2  &=& \alpha^2 B_2^{\rm Tree}
 \left[
1 - {9 g^2 (1+Y_2) \over 2 (4\pi f)^2} m_\pi^2\log{m_\pi^2\over
\mu^2}  + o_2(\mu)  m_\pi^2 \right]\,,\eea
or by recalling that
\bea
f_q = \alpha \left[ 1 - {1+3g^2\over (4\pi f)^2 } {3\over 4}m_\pi^2\log{m_\pi^2\over \mu^2}
+ c_f(\mu)  m_\pi^2  \right]\,,
\eea
for the bag parameters we have
\bea\label{bbbag}
B_1 &=& B_1^{\rm Tree} \left[ 1 + {1- 3g^2\over (4\pi f)^2 } {3\over 2}m_\pi^2\log{m_\pi^2\over \mu^2}
+ b_1(\mu)  m_\pi^2  \right]\,,\cr
B_2 &=& B_2^{\rm Tree} \left[ 1 +  {1- 3g^2 Y_2 \over (4\pi f)^2 } {3\over 2}m_\pi^2\log{m_\pi^2\over \mu^2}
+ b_2(\mu)  m_\pi^2  \right]\,,
\eea
where $g^2$ can be computed separately on the lattice as in ref.~\cite{glat}, and the parameters of the fit are 
$B_{1,2}^{\rm Tree}$ and the counterterms $b_{1,2}(\mu)$. It is worth emphasizing that the $\mu$-dependence 
in the chiral logarithms cancels against the one in the low energy constants. The situation with 
the chiral corrections to the  matrix element $O_{S-P}$ is similar to what we discussed in ref.~\cite{our} where 
for the non-Standard Model $\Delta B=2$ operators the new low energy constant ``$Y$" appeared. 
Its value is likely to be very close to unity as it represents the following ratio 
\bea\label{y2}
Y_2= {\langle B^\ast\vert O_{S-P}\vert B^\ast\rangle\over  \langle B \vert O_{S-P}\vert B \rangle }\;,
\eea
and it can be relatively easily evaluated on the lattice.~\footnote{Notice that the complexity related to 
the matching of the operator computed on the lattice to its counterpart renormalised in the continuum 
renormalisation scheme completely cancels in that ratio. }
Finally, let us stress once again that thanks to the identity
\bea
{1 \over 2} \lambda^A_{ab}  \lambda^A_{cd} = \delta_{ad}\delta_{bc} - {1\over 3} \delta_{ab}\delta_{cd}\,,
\eea
the chiral logarithms to the bag parameters $\varepsilon_{1,2}$ are of  the same as those in $B_{1,2}$ parameters but 
their low energy constants are of course different. To be fully explicit:
\bea
\varepsilon_1 &=& \varepsilon_1^{\rm Tree} \left[ 1 + {1- 3g^2\over (4\pi f)^2 } {3\over 2}m_\pi^2\log{m_\pi^2\over \mu^2}
+ b^\prime_1(\mu)  m_\pi^2  \right]\,,\cr
\varepsilon_2 &=& \varepsilon_2^{\rm Tree} \left[ 1 +  {1- 3g^2 Y^\prime_2 \over (4\pi f)^2 } {3\over 2}m_\pi^2\log{m_\pi^2\over \mu^2}
+ b^\prime_2(\mu)  m_\pi^2  \right]\,,
\eea
with 
\bea\label{y2prime}
Y_2^\prime= {\langle B^\ast\vert T_{S-P}\vert B^\ast\rangle\over  \langle B \vert T_{S-P}\vert B \rangle }\;.
\eea

\noindent{\bf 3.2. Back to phenomenology}:
In the early phenomenological applications  the formulas derived in HMChPT 
were used to estimate the size  of the hadronic quantities by using the theory with $N_f=3$ light flavour and 
by neglecting the counterterms (or, at best,  estimating them by
means of some quark model). Nowadays we also know that the $\left(\frac{1}{2}\right)^+$-states should be included if one is to use HMChPT  with $N_f=3$. In what follows, the $\left(\frac{1}{2}\right)^+$-contributions  will be neglected too, 
which is an extra assumption. 
To get the difference $B_2-B_1$ we will  proceed along these lines and 
impose $B_{1,2}^{\rm Tree}=1$, like in the vacuum saturation approximation,  and 
neglect the counterterms, to obtain
\bea
B_2-B_1 &=&
{3 g^2 (1-Y_2) \over (4\pi f)^2} \left( \frac{3}{2}
m_\pi^2\log{m_\pi^2\over \mu^2} + m_K^2\log{m_K^2\over \mu^2} +
\frac{1}{6} m_\eta^2\log{m_\eta^2\over \mu^2} \right) \,,
 \eea
which for $g^2\approx 0.3$, $f=120$~MeV, and $\mu=1$~GeV gives
\bea
B_2-B_1 = 0.21\  (1-Y_2)\,.
\eea
This is as far as one can get at this stage, since there is no  information available concerning the size of $Y_2$. 
We reiterate that it  can be computed on the lattice as indicated in eq.~(\ref{y2}).  Note in passing that if we use $B_2-B_1=0.05$ as inferred in introduction from the $D$-decays, we would obtain 
$Y_2\approx 0.8$. 
Similarly, for the bag parameters entering eq.~(\ref{lifediff}) we have
\bea
B_1^q&=& 1+ {1- 3 g^2 \over (4\pi f)^2}  \left[\frac{3}{2} m_\pi^2\log{m_\pi^2 } + m_K^2\log{m_K^2 } +
\frac{1}{6} m_\eta^2\log{m_\eta^2} \right] = 0.98\,,\cr
B_1^s&=& 1+ {1- 3 g^2 \over (4\pi f)^2}  \left[2 m_K^2\log{m_K^2 } +
\frac{2}{3} m_\eta^2\log{m_\eta^2} \right] = 0.96\,,
\eea
while for
\bea
B_2^q= 0.77 + 0.21\ Y_2 \,,\qquad B_2^s= 0.59 + 0.37\ Y_2\,.
\eea
which, together with the assumption that $\varepsilon_{1,2}^{\rm Tree}=\varepsilon_{1,2}^{\rm VSA}=0$ 
brings eq.~(\ref{lifediff}) to
\bea
{\tau(B^\pm)\over \tau(B_d)} &=& 1.077+0.002\ Y_2\qquad \left( 1.076\pm 0.008\right)^{\rm exp}  \,,\nn\\
{\tau(B_s)\over \tau(B_d)}  &=& 1.001- 0.002\ Y_2 \qquad \left( 0.950\pm 0.019\right)^{\rm exp} \,.
\eea
where in the parentheses we also give the experimental values~\cite{PDG}.  We see that in spite of the assumptions 
the current experimental information does not allow to constrain appreciably the value of the coupling $Y_2$.

\vspace*{.75cm}

\noindent\underline{\bf 4. Summary}:
In this paper we presented the result of our calculation of the chiral corrections to the matrix elements of four-quark 
$\Delta B=0$ operators that are relevant to the phenomenology of the lifetime ratios of the heavy-light mesons 
and to the inclusive semileptonic decay spectra~\cite{lenz2}. The calculation of the chiral corrections can be combined with the lattice calculations of the $\Delta B=0$ matrix elements to either extrapolate the lattice data towards the physical light quark masses, and/or to fix the counterterm coefficients $b_{1,2}(\mu)$ in eq.~(\ref{bbbag}), the couplings  $Y_2$~(\ref{y2})
and $Y_2^\prime$~(\ref{y2prime}) and the tree level bag parameters (in terms of chiral expansion).

\vskip 12mm

\noindent\underline{\bf Acknowledgement}: 
This work has been supported by the EU-RTN Programme,   Contract No. MRTN--CT-2006-035482, ``Flavianet'', 
the Slovenian research Agency and the ANR contract ``DIAM'' JC07\_204962.

\end{document}